\documentclass[letterpaper]{article} 
\usepackage{aaai2026}  
\nocopyright
\usepackage{times}  
\usepackage{helvet}  
\usepackage{courier}  
\usepackage[hyphens]{url}  
\usepackage{graphicx} 
\usepackage{natbib}  
\usepackage{caption} 
\usepackage{booktabs}
\usepackage{amsmath}
\usepackage{amssymb}
\usepackage{multirow}
\usepackage{subfig}
\usepackage[table,xcdraw]{xcolor}
\usepackage{colortbl}
\usepackage{pifont}
\usepackage{algorithm}
\usepackage{algorithmic}

\usepackage{newfloat}
\usepackage{listings}
\DeclareCaptionStyle{ruled}{labelfont=normalfont,labelsep=colon,strut=off} 
\floatstyle{ruled}
\newfloat{listing}{tb}{lst}{}
\floatname{listing}{Listing}
\title{TuningIQA: Fine-Grained Blind Image Quality Assessment for \\ Livestreaming Camera Tuning}
\author{
    Xiangfei Sheng\textsuperscript{\rm 1}\equalcontrib,
    Zhichao Duan\textsuperscript{\rm 1}\equalcontrib,
    Xiaofeng Pan\textsuperscript{\rm 1},
    Yipo Huang\textsuperscript{\rm 2},
    Zhichao Yang\textsuperscript{\rm 1},
    Pengfei Chen\textsuperscript{\rm 1},
    Leida Li\textsuperscript{\rm 1,3}\thanks{Corresponding author}
}
\affiliations{
    \textsuperscript{\rm 1}School of Artificial Intelligence, Xidian University,\textsuperscript{\rm 2}School of Data Science and Artificial Intelligence, Chang'an University, \textsuperscript{\rm 3}State Key Laboratory of Electromechanical Integrated Manufacturing of High-Performance Electronic Equipments, \\Xidian University\\
    xiangfeisheng@gmail.com, zach@stu.xidian.edu.cn, panxf@stu.xidian.edu.cn, yphuang@chd.edu.cn, yangzhichao@stu.xidian.edu.cn, chenpengfei@stu.xidian.edu.cn, ldli@xidian.edu.cn
}

\usepackage{bibentry}

\begin{document}

\maketitle
\vspace*{-3em} 
\begin{abstract}
Livestreaming has become increasingly prevalent in modern visual communication, where automatic camera quality tuning is essential for delivering superior user Quality of Experience (QoE). Such tuning requires accurate blind image quality assessment (BIQA) to guide parameter optimization decisions. Unfortunately, the existing BIQA models typically only predict an overall coarse-grained quality score, which cannot provide fine-grained perceptual guidance for precise camera parameter tuning. To bridge this gap, we first establish \textbf{FGLive-10K}, a comprehensive fine-grained BIQA database containing 10,185 high-resolution images captured under varying camera parameter configurations across diverse livestreaming scenarios. The dataset features 50,925 multi-attribute quality annotations and 19,234 fine-grained pairwise preference annotations. Based on FGLive-10K, we further develop \textbf{TuningIQA}, a fine-grained BIQA metric for livestreaming camera tuning, which integrates human-aware feature extraction and graph-based camera parameter fusion. Extensive experiments and comparisons demonstrate that TuningIQA significantly outperforms state-of-the-art BIQA methods in both score regression and fine-grained quality ranking, achieving superior performance when deployed for livestreaming camera tuning.
\end{abstract}

\section{Introduction}

With the rapid expansion of mobile livestreaming across entertainment and e-commerce platforms (\textit{e.g.}, TikTok Live, YouTube Shorts), automatic camera parameter tuning has become increasingly important for delivering superior Quality of Experience (QoE) \cite{10159883, streaming}. Unlike traditional photography where post-capture editing can compensate for suboptimal settings, livestreaming demands real-time optimization during capture, making automated camera tuning essential for both professional streamers and casual users lacking photography expertise.

\begin{figure}[t]
  \centering
  \includegraphics[width=0.98\linewidth]{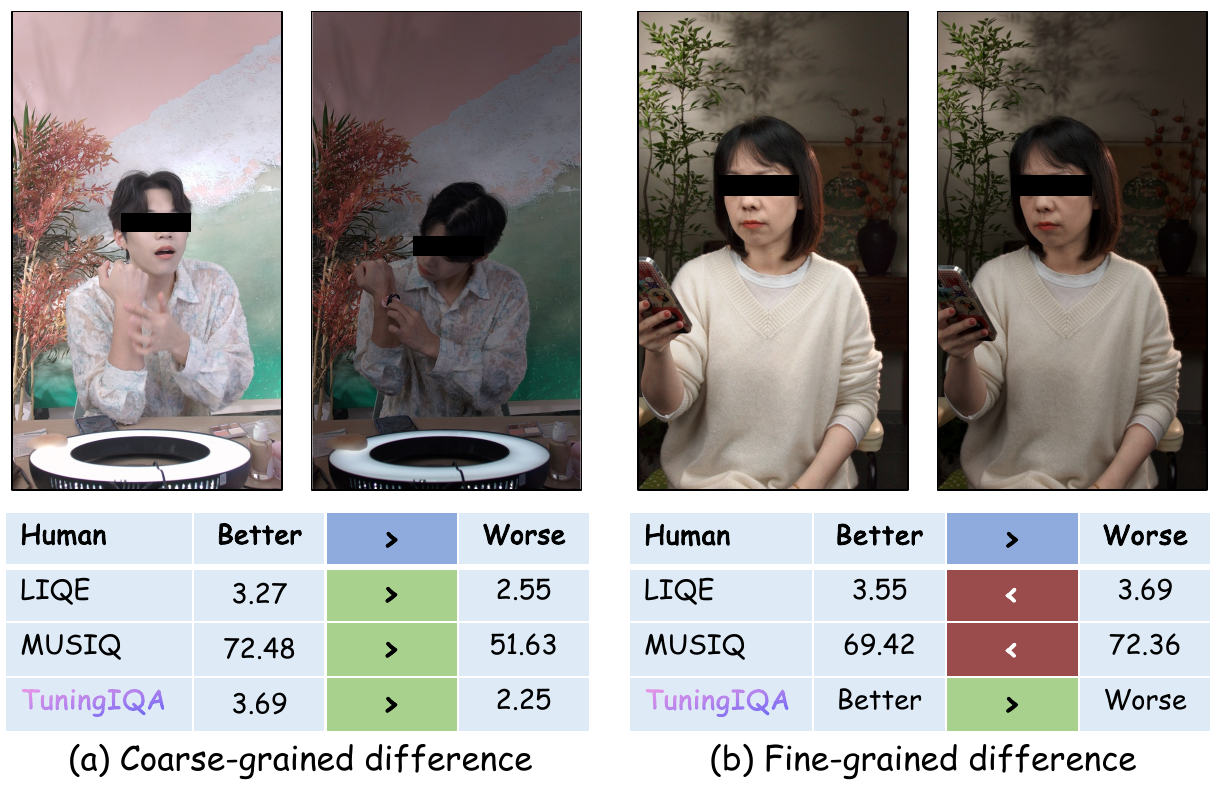}
  \caption{Illustration of the fine-grained challenge in IQA. Coarse-grained quality difference (a) and fine-grained quality difference (b) caused by inappropriate camera parameters. State-of-the-art score-based methods LIQE \cite{LIQE} and MUSIQ \cite{MUSIQ} fail to capture fine-grained quality differences.}
  \label{fig:teaser}
\end{figure}

\begin{figure*}[!t]
  \centering
  \includegraphics[width=0.95\linewidth]{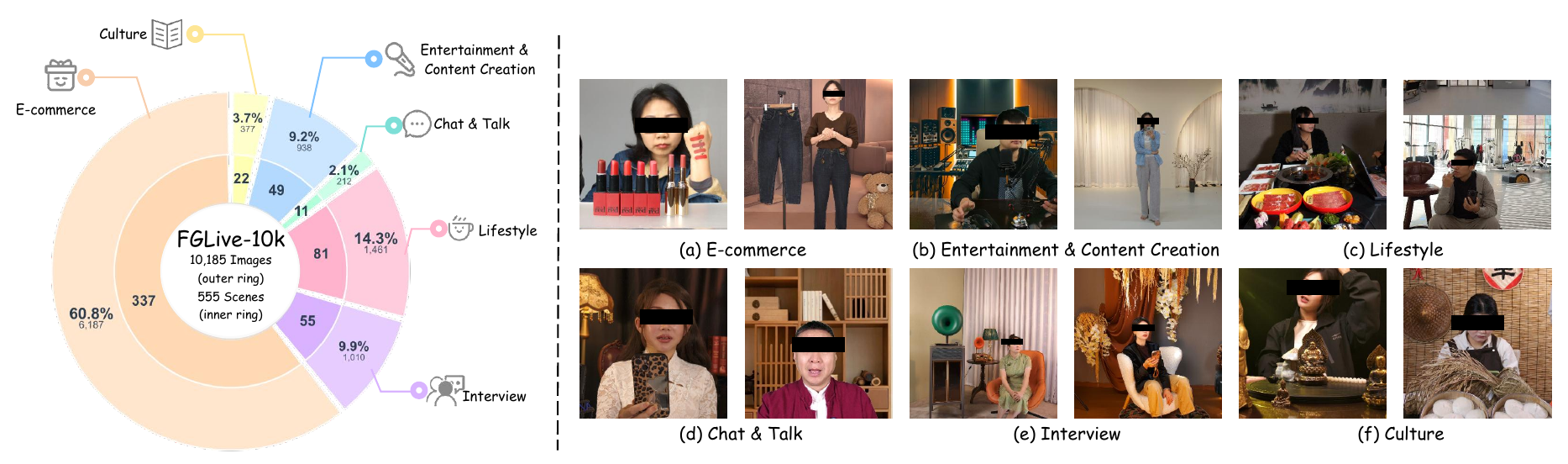}
  \caption{Statistics of the FGLive-10K dataset and representative samples covering diverse livestreaming scenarios.}
  \label{fig:fglive_sample}
\end{figure*}

Image quality assessment (IQA) constitutes the core of effective camera tuning. In practice, automated camera tuning follows a common approach: algorithms perform parameter adjustments within consistent scenes, followed by an objective metric to predict the output image quality and provide feedback to the parameter tuning, producing the best quality image. 
Therefore, the fundamental challenge lies in precise blind image quality assessment (BIQA), particularly during the fine-grained parameter optimization process where quality variations are often subtle and may not dramatically change overall image appearance. In these scenarios, fine-grained quality assessment becomes essential for distinguishing between parameter configurations that produce visually similar but still \textit{noticeable} perceptual differences.                                              

State-of-the-art BIQA methods \cite{LIQE, MUSIQ, CLIPIQA} designed for score-based evaluation, face fundamental limitations when applied to camera tuning scenarios. As illustrated in Figure \ref{fig:teaser}, while these methods demonstrate accuracy for coarse-grained quality discrimination, they struggle with fine-grained assessment, which is essential for parameter optimization. When parameter adjustments produce subtle quality changes, existing BIQA methods often fail to provide reliable guidance, restricting their practical utility in automated camera tuning systems.

Livestreaming camera tuning also presents two specific requirements, which are inadequately addressed by general-purpose BIQA methods: (i) \textbf{Human-region Priority}, emphasizing image quality evaluation on human-centric regions such as faces and bodies, reflecting the heightened sensitivity of the human visual system (HVS) \cite{HVS-BOOK} towards these areas over background elements. (ii) \textbf{Parameter Interdependency Modeling}, capturing complex physical relationships among camera settings that jointly influence image quality. Camera parameters exhibit intricate interdependencies rooted in photographic principles—adjusting one parameter often necessitates compensating changes in others to maintain desired quality. Current BIQA methods fail to model these parameter interactions and lack the specialized architectural designs needed for camera-guided quality optimization.

To address the above limitations, we present multi-fold technical contributions spanning dataset construction, method development, and experimental analysis:

$\bullet$ \textbf{Dataset.} We construct FGLive-10K, the first-of-its-kind fine-grained BIQA dataset specifically designed for livestreaming camera tuning. As shown in Figure \ref{fig:fglive_sample}, the dataset comprises 10,185 high-resolution images from 555 distinct scenes, each containing systematic camera parameter variations. Beyond conventional Mean Opinion Scores (MOS), we provide fine-grained pairwise preference annotations, enabling reliable discrimination of subtle quality differences essential for parameter optimization.

$\bullet$ \textbf{Method.} We propose TuningIQA, a fine-grained BIQA framework with two key innovations. First, we design a Human-aware Feature Extraction (HFE) module that prioritizes human-centric regions through aesthetics-guided feature learning. Second, we introduce an optional Graph-based Camera Parameter Fusion (GCPF) module that models physical relationships among camera settings through graph attention networks. The framework unifies multi-attribute regression and fine-grained rank learning for comprehensive quality assessment and precise camera tuning guidance.

$\bullet$ \textbf{Experimental Analysis.} Extensive experiments demonstrate TuningIQA's superior performance over state-of-the-art methods in both coarse-grained quality score prediction and fine-grained quality ranking. Our analysis reveals fundamental limitations in existing score-based approaches for fine-grained assessment, while practical camera tuning experiments show 74-76\% win rates against leading BIQA methods, confirming the effectiveness of our specialized design for camera tuning applications.

\begin{table}[!t]
\centering
\caption{Comparison of popular BIQA datasets with authentic distortions.}
\label{tab:databases}
\small  
\renewcommand{\arraystretch}{1.0}
\begin{tabular*}{\linewidth}{@{\extracolsep{\fill}}lcccl@{}}
\toprule
\textbf{Dataset} & \textbf{\#Img} & \textbf{\#Param} & \textbf{\#Attr} & \textbf{Annotation} \\
\midrule
BID & 585 & 0 & 0 & MOS \\
CID2013 & 480 & 0 & 4 & MOS \\
LIVE Challenge & 1,162 & 0 & 0 & MOS \\
KonIQ-10K & 10,073 & 0 & 4 & MOS \\
SPAQ & 11,125 & 0 & 5 & MOS \\
\midrule
\rowcolor[HTML]{F5F5F5}
\textbf{FGLive-10K} & 10,185 & 7 & 4 & MOS+Rank \\
\bottomrule
\end{tabular*}
\end{table}

\section{Related Work}

\subsection{IQA Benchmarks}

Benchmark datasets have driven significant progress in image quality assessment research. Early datasets focused on synthetic distortions like LIVE \cite{LIVE}, TID2013 \cite{TID2013}, and CSIQ \cite{CSIQ}, using simulated degradations on reference images to enable full-reference IQA development. However, obtaining pristine references proves impractical in real scenarios, motivating research toward authentic distortions. Pioneering efforts include BID \cite{RBID} with 585 DSLR-captured blur images, LIVE Challenge \cite{LIVEW} featuring 1,162 mobile-captured images with crowdsourced annotations, and KonIQ-10k \cite{KonIQ-10k} expanding diversity with 10,000 multimedia images. SPAQ \cite{SPAQ} incorporated EXIF metadata and laboratory-controlled annotations for smartphone imaging scenarios. A brief comparison of authentic BIQA datasets is summarized in Table \ref{tab:databases}.

\subsection{IQA Method}  

Early IQA methods relied on statistical priors \cite{ILNIQE, NIQE} and handcrafted natural scene statistics \cite{BRISQUE, CORNIA, BLIINDS-II}, achieving reasonable success on synthetic distortions but struggling with authentic degradations \cite{LIVEW}. The shift toward real-world assessment drove CNN-based approaches \cite{DeepIQA, DB-CNN, WaDIQaM-NR} that learned hierarchical quality representations directly from data. Recent advances include architectural innovations like HyperIQA's semantic-quality disentanglement \cite{HyperNet} and Transformer-based models with attention mechanisms \cite{LIQE, MUSIQ, CLIPIQA, VT-IQA, MANIQA}. More recently, MLLM-based methods \cite{Q-Align, deqa_score} have emerged, demonstrating superior generalization capabilities and interpretability.

Despite considerable advancements, existing methods exhibit fundamental limitations when applied to livestreaming camera tuning. (1) Most methods target general-purpose quality evaluation, overlooking key characteristics of livestreaming environments such as human-centric visual focus and systematic camera parameter variations. (2) Existing methods typically produce single coarse-grained quality scores, which are inherently limited in identifying subtle differences essential for fine-grained perception. These limitations underscore the need for a specialized fine-grained BIQA framework tailored for livestreaming camera tuning, motivating the development of TuningIQA.

\begin{figure}[!t]
  \centering
  \includegraphics[width=\linewidth]{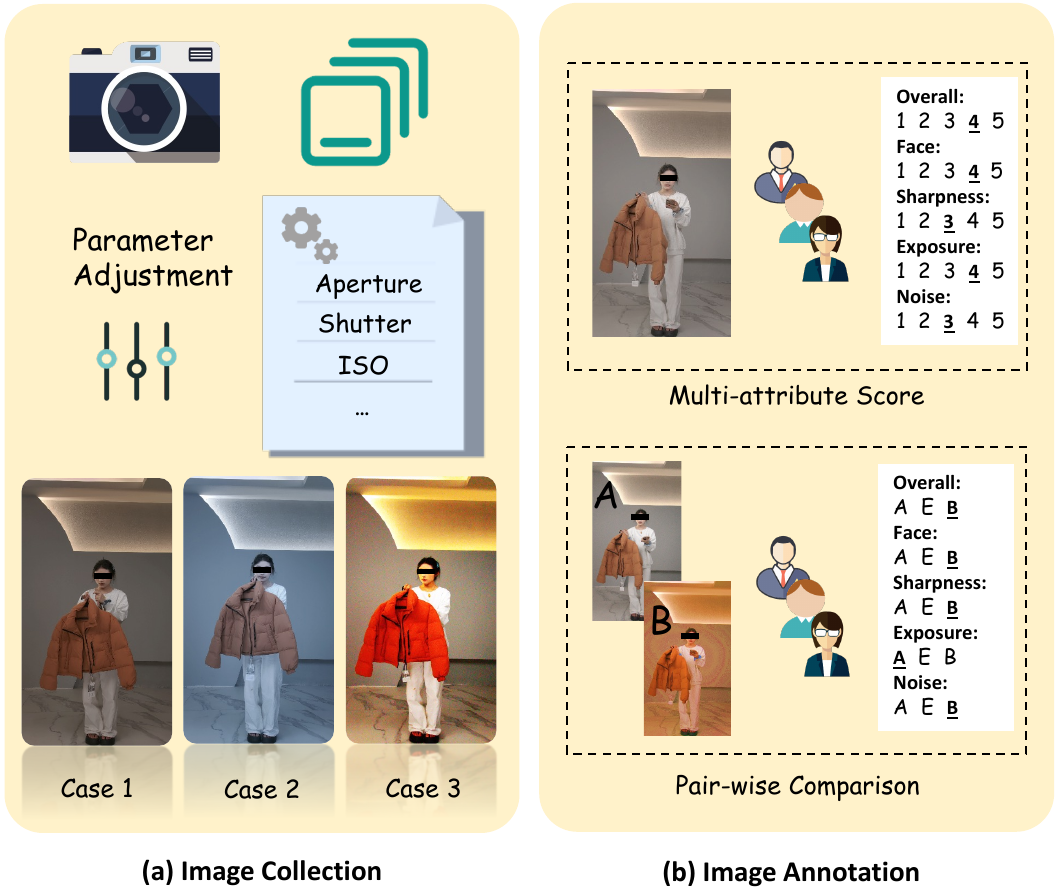}
  \caption{Details of the FGLive-10K dataset construction. (a) Various distorted images are collected by adjusting the camera parameters. (b) Multi-attribute quality score annotation and pairwise comparison of fine-grained image pair.}
  \label{fig:data_main}
\end{figure}

\section{FGLive-10K}
\subsection{Dataset Construction}

We describe the image collection and annotation processes for FGLive-10K, designed specifically for livestreaming camera tuning scenarios. It should be noted that all depicted individuals provided consent for use in this research.

\subsubsection{Image Collection.}

To capture camera parameter-induced distortions, we employed three internally developed livestreaming cameras in authentic usage scenarios. Field Application Engineers first optimized reference parameters for each scenario, then generated distorted images by introducing systematic deviations to aperture, shutter speed, ISO, white balance, contrast, saturation, and sharpness. Figure \ref{fig:data_main}(a) demonstrates parameter-induced variations within a single scene. The collected images span typical livestreaming applications including e-commerce, entertainment, and creative content. After filtering images without human subjects, FGLive-10K comprises 10,185 high-resolution (1920×1080) images from 555 distinct scenes.

To explore Parameter Interdependency Modeling, we preserved the complete 7-parameter metadata for a subset of images, creating \textbf{FineLive-p} with 5,559 training images and 1,148 test images. This subset follows the same annotation protocol as the main dataset.

\subsubsection{Image Annotation.}

We invited 25 volunteers with diverse backgrounds (IQA researchers, photography enthusiasts, art students) to participate in subjective experiments following ITU-R BT.500-15 recommendations \cite{series2012methodology}.

\textbf{Multi-attribute MOS.} Annotators assigned discrete scores (1-5: bad to excellent) for overall quality and four critical attributes: \textit{face quality} (visual quality of human facial regions), \textit{sharpness} (image clarity and detail preservation), \textit{exposure} (brightness and contrast appropriateness), and \textit{noise} (absence of visual artifacts and grain). After post-screening based on Pearson correlation, each image received at least 16 annotations, yielding Mean Opinion Scores:
\begin{equation}
s^{\text{attr}} = \frac{1}{N} \sum_{i=1}^N s_i^{\text{attr}},
\end{equation}
where $s^{\text{attr}}$ denotes the MOS for a specific attribute, $N$ is the number of annotators.

\textbf{Pairwise Comparison.} To address the limitation of MOS in fine-grained quality discrimination, we constructed image pairs with subtle differences within each scene and conducted pairwise comparisons. Initial preferences derived from MOS comparisons ($c_{pq}^* = \mathbb{I}(s_p > s_q)$) were refined for pairs with $\Delta s = |s_p - s_q| \leq 0.8$, as these exhibit significant annotation uncertainty \cite{katsigiannis2018interpreting}. Fine-grained pairs underwent human verification where annotators selected preferred images or marked equivalence. The final preference was:
\begin{equation}
c_{pq} =
\begin{cases}
c_{pq}^*, & \text{if } \Delta s > 0.8, \\
\frac{1}{K}\sum_{k=1}^K \psi_k(I_p, I_q), & \text{if } \Delta s \leq 0.8,
\end{cases}
\end{equation}
where $K$ denotes annotations per pair and $\psi_k \in \{0,0.5,1\}$ represents individual judgments (worse, equivalent, better quality of $I_p$ vs $I_q$).
This strategy produced 91,946 annotated pairs, with 21\% (19,234 pairs) refined through pairwise comparison.

\begin{figure}[!t]
  \centering
  \includegraphics[width=0.95\linewidth]{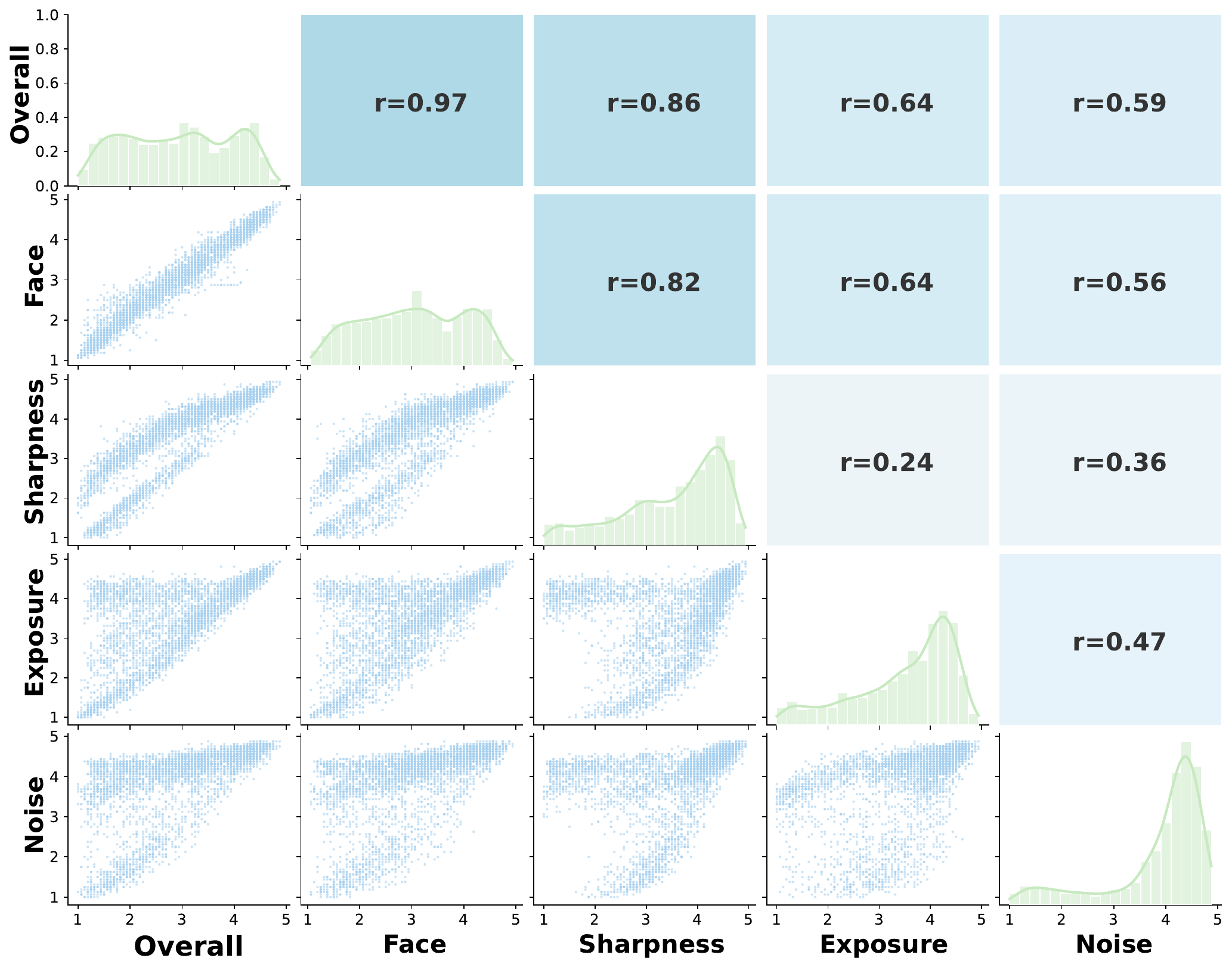}
  \caption{Relationships among quality dimensions: scatterplot (bottom left), Pearson correlation (top right), annotation distribution (diagonal).}
  \label{fig:mos_dist}
\end{figure}

\begin{figure}[!t]
  \centering
  \includegraphics[width=0.9\linewidth]{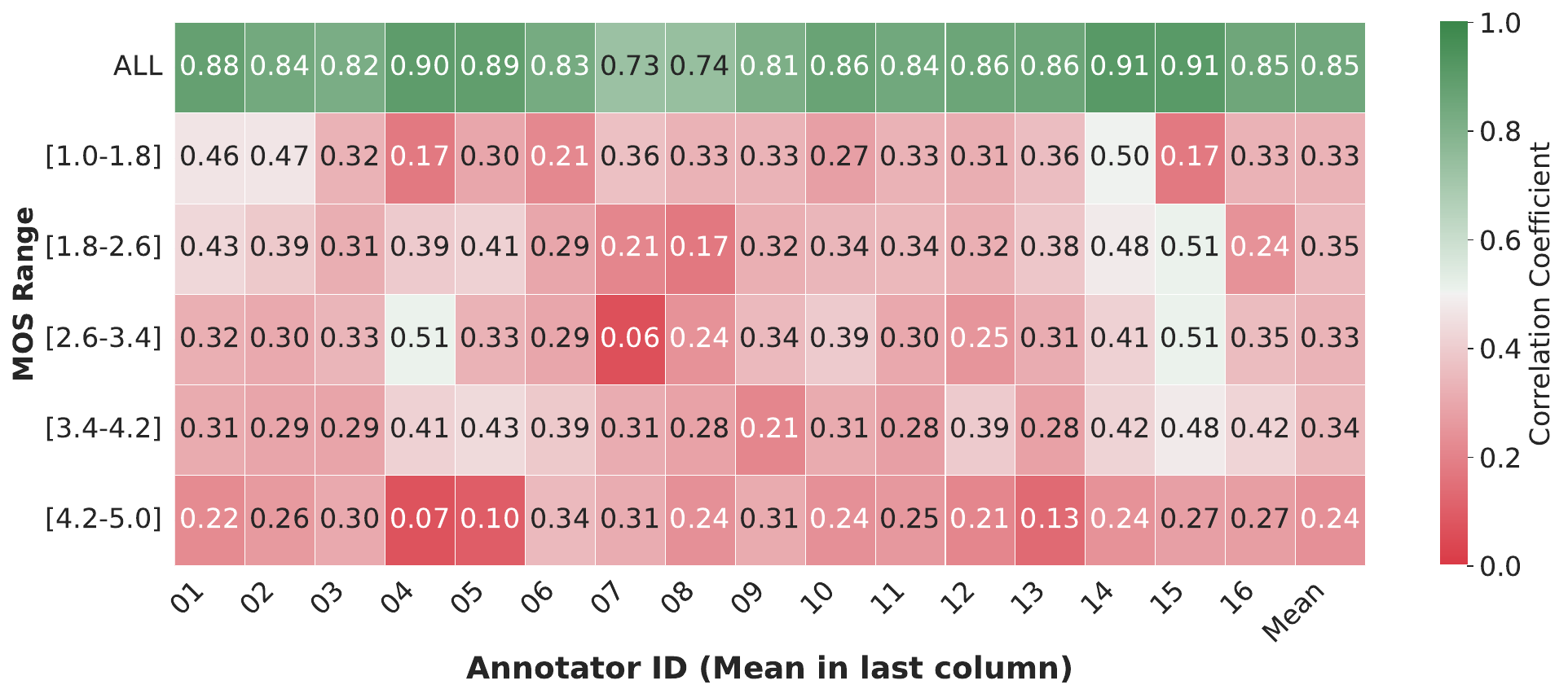}
  \caption{Correlation between Individual Annotation and MOS within specific quality tiers (MOS range).}
  \label{fig:annotator_analysis}
\end{figure}


\subsection{Statistics and Analysis}

FGLive-10K contains 555 scenes with 10,185 images annotated across five quality attributes. The dataset is split into 451 training scenes (8,148 images, 72,843 pairs) and 104 test scenes (2,037 images, 3,705 fine-grained pairs).

\textbf{Annotation Distribution.} The FGLive-10K dataset features comprehensive multi-attribute quality annotations, where distributions are visualized along the diagonal in Figure \ref{fig:mos_dist}. The inter-attribute variations reveal their distinct perceptual value. The uniform distribution of overall quality scores confirms our dataset’s balanced coverage of diverse quality levels across various scenarios.

\textbf{Attribute Correlation.} 
To study FGLive-10K from the correlation perspective, we visualize the Pearson Linear Correlation Coefficient (PLCC) and scatterplot among each dimension in Figure \ref{fig:mos_dist}. It can be observed that the correlation between overall and face is extremely high, which indicates that face quality can largely affect overall perception. Low correlations among sharpness, exposure, and noise dimensions suggest their complementary yet independent contributions. This divergence highlights the necessity of multi-attribute evaluation for precise distortion perception.

\textbf{Individual Annotation Consistency.} We investigate annotator reliability through correlation analysis in Figure \ref{fig:annotator_analysis}. Post-screened annotators strongly agree with MOS (average PLCC=0.85), confirming annotation reliability at the coarse-grain level. However, when analyzing within smaller MOS tiers (\textit{e.g.} Figure \ref{fig:annotator_analysis} rows 2-6), the correlations drop significantly (\textit{i.e.} PLCC=0.24 in the 4.2-5.0 tier). Which highlights single-stimulus limitations in discriminating subtle quality differences. Such findings validate the necessity of pairwise comparison refinement for fine-grained image pairs.

\begin{figure*}[!t]
  \centering
  \includegraphics[width=0.9\linewidth]{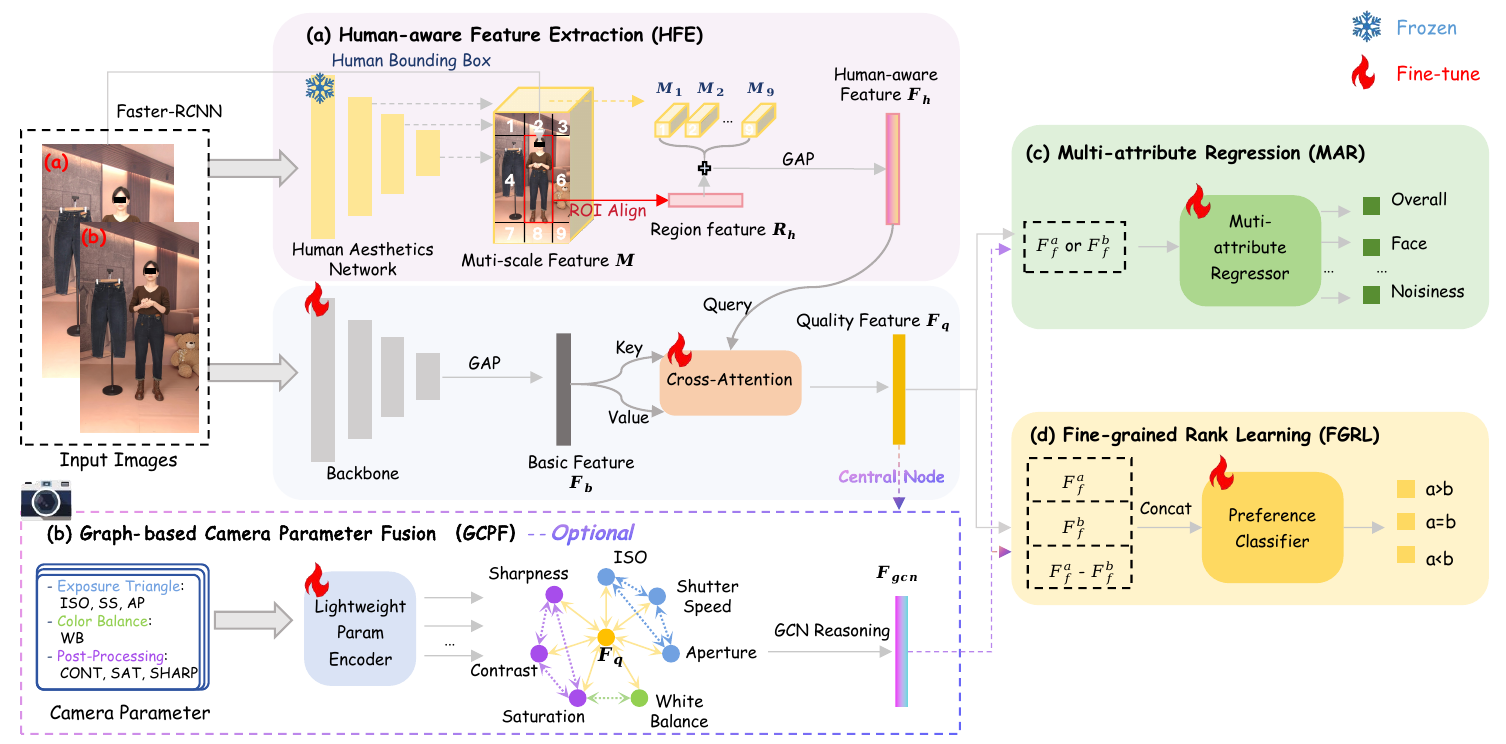}
  \caption{Overview of the proposed TuningIQA framework.}
  \label{fig:method}
\end{figure*}

\section{TuningIQA Metric}

Based on insights from the FGLive-10K dataset, we further propose TuningIQA, a fine-grained quality assessment framework for livestreaming camera tuning. The framework supports two operational modes: \textit{single-image mode} for direct quality scoring and \textit{pairwise mode} for fine-grained comparison. The pipeline is illustrated in Figure \ref{fig:method}.
TuningIQA is built upon two core modules: \textit{human-centric feature extraction} which prioritizes perceptually critical regions, and \textit{graph-based camera parameter fusion} that models camera setting interdependencies.

\subsection{Human-aware Feature Extraction}

To address the human-region priority, we design a Human-aware Feature Extraction (HFE) module that emphasizes human-centric regions. Specifically, we first employ Faster R-CNN \cite{ren2016faster} to detect human subjects in input images, obtaining human bounding boxes $\text{BBox}_h$ that guide subsequent feature extraction. To understand human-centric visual quality, we leverage a Human Aesthetics Network built upon the EfficientNetV2-M \cite{tan2021efficientnetv2} backbone, pre-trained on the human subset of the TAD66K dataset \cite{TANet}. This backbone extracts a multi-scale feature map $\mathbf{M}$ with $C$ channels that captures aesthetic-aware representations.

Given the detected human bounding boxes, we extract human region features $\mathbf{R}^h$ using ROI Align operations. The dimension of $\mathbf{R}^h$ is reduced to $\frac{C}{2}$ and then appended to each spatial location of $\mathbf{M}$, creating an enhanced feature representation that integrates both global image context and human-centric information. We then partition the enhanced feature map into nine spatial regions $\{\hat{\mathbf{M}}_k\}_{k=1}^9$ centered around the human subjects. To explicitly handle different partitions, we employ nine individual nonlinear transformations with residual learning:
\begin{equation}
\mathbf{F}_k = \Phi_k(\hat{\mathbf{M}}_k) + \hat{\mathbf{M}}_k, \quad k \in \{1, 2, \ldots, 9\},
\end{equation}
where $\Phi_k(\cdot)$ represents partition-specific transformations implemented as $3 \times 3$ convolutional layers with $C$ output channels.

The updated partitions $\{\mathbf{F}_k\}_{k=1}^9$ are combined to form the human-aware feature map. Then we employ a cross-attention mechanism that fuses human-aware features $\mathbf{F}_h$ (obtained through Global Average Pooling applied to the combined partitions) with basic backbone features $\mathbf{F}_b$:
\begin{equation}
\mathbf{F}_q = \text{CrossAttention}(\mathbf{Q}, \mathbf{K}, \mathbf{V}) = \text{softmax}\left(\frac{\mathbf{Q}\mathbf{K}^T}{\sqrt{d_k}}\right)\mathbf{V}
\end{equation},
where $\mathbf{Q} = \mathbf{F}_h \mathbf{W}_Q$, $\mathbf{K} = \mathbf{F}_b \mathbf{W}_K$, $\mathbf{V} = \mathbf{F}_b \mathbf{W}_V$ are learned projections. The resulting quality-aware features $\mathbf{F}_q$ effectively integrate human-centric visual cues with global image context for comprehensive quality assessment.

\begin{table*}[]
\centering
\caption{Performance of TuningIQA and the state-of-the-art BIQA methods in two tasks: quality score regression and fine-grained ranking on the FGLive-10K database.}
\label{tab:main}
\small  
\renewcommand{\arraystretch}{1.0}
\setlength{\tabcolsep}{6pt}  
\begin{tabular*}{\textwidth}{@{\extracolsep{\fill}}lllccccc@{}}
\toprule
& & & & \multicolumn{2}{c}{\textbf{Score Regression}} & \multicolumn{1}{c}{\textbf{FG Ranking}} \\
\cmidrule(lr){5-6} \cmidrule(lr){7-7}
\multirow{-2}{*}{\textbf{Method}} & \multirow{-2}{*}{\textbf{Backbone}} & \multirow{-2}{*}{\textbf{\#Param}} & \multirow{-2}{*}{\textbf{Input Size}} & \textbf{SRCC↑} & \textbf{PLCC↑} & \textbf{FG-ACC↑} \\ 
\midrule
NIQE \cite{NIQE}         & --                & --      & Full res. & 0.4512 & 0.5065 & 0.4327 \\
ILNIQE \cite{ILNIQE}     & --                & --      & Full res. & 0.5706 & 0.5412 & 0.4089 \\
BRISQUE \cite{BRISQUE}   & --                & --      & Full res. & 0.3411 & 0.4280  & 0.4583 \\
\midrule
DBCNN \cite{DB-CNN}      & VGG16             & 15.31M  & 224×224   & 0.7507 & 0.7495 & 0.5667 \\
HyperIQA \cite{HyperNet} & ResNet-50         & 27.38M  & 224×224   & 0.8472 & 0.8518 & 0.5980 \\
MT-A \cite{SPAQ}         & ResNet-50         & 23.57M  & 512×512   & 0.8821 & 0.8829 & 0.6223 \\
CLIP-IQA+ \cite{CLIPIQA} & CLIP (ResNet-50)   & 149.70M & Full res. & 0.7056 & 0.7039 & 0.5570 \\
MUSIQ \cite{MUSIQ}       & Transformer       & 157.23M & Full res. & 0.8662 & 0.8687 & 0.6861 \\
SARQUE \cite{SARQUE}     & MobileNetV3-S     & 8.85M   & 224×224   & 0.8581 & 0.8591 & 0.6095 \\
LIQE \cite{LIQE}         & CLIP (ViT-B/32)    & 151.28M & 224×224   & 0.9235 & 0.9206 & 0.6533 \\
Q-Align \cite{Q-Align}         & mPLUG-Owl2-7B    & 8197.86M & 448×448   & 0.8721 & 0.8315 & 0.6311 \\
Compare2Score \cite{zhu2024adaptive}         & mPLUG-Owl2-7B    & 8197.86M & 448×448   & 0.8502 & 0.8190 & 0.6266 \\

\midrule
\rowcolor[HTML]{EFEFEF}
TuningIQA                & MobileNetV3-S     & 58.29M   & 224×224   & \underline{0.9308} & \underline{0.9302} & \underline{0.7065} \\
\rowcolor[HTML]{EFEFEF}
TuningIQA                & ResNet-50          & 84.62M  & 224×224   & \textbf{0.9385} & \textbf{0.9364} & \textbf{0.7284} \\ 
\bottomrule
\end{tabular*}
\end{table*}

\subsection{Graph-based Camera Parameter Fusion}
For images with available camera metadata, we introduce an \textit{optional} Graph-based Camera Parameter Fusion (GCPF) module that explicitly models these parameter interactions through graph neural networks.

\noindent\textbf{Graph Construction and Node Encoding.} We construct a heterogeneous graph $\mathcal{G} = (\mathcal{V}, \mathcal{E})$ where nodes $\mathcal{V}$ represent both visual features and camera parameters. The visual node serves as the central hub that aggregates information from all parameter nodes, enabling rich cross-modal interactions. Specifically, we define eight nodes: one visual node encoding $\mathbf{F}_q$ and seven parameter nodes. Each parameter node is encoded through individual linear transformations:
\begin{equation}
\mathbf{v}_{\text{visual}} = \mathbf{W}_v \mathbf{F}_q, \quad \mathbf{v}_i = \mathbf{W}_p p_i,
\end{equation}
where $\mathbf{W}_v$ and $\mathbf{W}_p$ are learned projection matrices, and $p_i$ represents the $i$-th camera parameter value.

\noindent\textbf{Physical Relationship Modeling.} The edge set $\mathcal{E}$ encodes both cross-modal connections and physical relationships based on photographic principles. We establish connections following relationships: (1) \textit{Cross-modal connections}: the visual node connects to all parameter nodes, enabling parameter-visual feature interaction; (2) \textit{Exposure Triangle}: ISO, shutter speed, and aperture form a strongly connected subgraph as they jointly determine exposure—increasing ISO compensates for faster shutter or smaller aperture; (3) \textit{Post-processing Chain}: contrast, saturation, and sharpness exhibit mutual dependencies in image enhancement, where higher contrast often requires adjusted saturation; (4) \textit{Color Correlation}: white balance directly influences saturation perception due to color temperature effects.

\noindent\textbf{Graph Reasoning.} The graph reasoning process employs two-layer Graph Attention Networks (GAT) \cite{velickovic2017graph} to propagate information across the parameter-visual graph:
\begin{equation}
\mathbf{H}^{(1)} = \text{GAT}_1(\mathbf{X}, \mathcal{E}), \quad \mathbf{H}^{(2)} = \text{GAT}_2(\mathbf{H}^{(1)}, \mathcal{E}),
\end{equation}
where $\mathbf{X} = [\mathbf{v}_{\text{visual}}, \mathbf{v}_1, \ldots, \mathbf{v}_7]$ contains all encoded node features. The first GAT layer employs 4 attention heads with concatenation to capture diverse relationship patterns, while the second layer uses single-head attention for feature integration. The final parameter-aware representation $\mathbf{F}_{GCN}$ is extracted from the updated central visual node:
\begin{equation}
\mathbf{F}_{gcn} = \mathbf{H}^{(2)}[0, :].
\end{equation}
where $\mathbf{H}^{(2)}[0, :]$ denotes the first row of matrix $\mathbf{H}^{(2)}$.
This approach enables the model to reason about parameter interactions and their combined effects on image quality, providing richer representations for subsequent quality prediction.
\subsection{Multi-attribute Regression and Fine-grained Ranking}

\noindent\textbf{Multi-attribute Regressor.} For comprehensive quality assessment, we implement five specialized prediction heads that estimate quality scores across different perceptual dimensions:
\begin{equation}
    \hat{s}^{\text{attr}} = \text{MLP}_{\theta^{\text{attr}}_s}(\mathbf{F}_{f}),
\end{equation}
where $\mathbf{F}_{f}$ represents the fused features (either $\mathbf{F}_q$ or $\mathbf{F}_{gcn}$ when parameters are available), and $\text{attr} \in \{\text{overall}, \text{face}, \text{sharpness}, \text{exposure}, \text{noise}\}$.

\noindent\textbf{Fine-grained Preference Classifier.} To capture subtle quality differences, we incorporate a pairwise comparison mechanism:
\begin{equation}
    \hat{c}^{\text{attr}} = \sigma\left(\text{MLP}_{\theta^{\text{attr}}_c}([\mathbf{F}_a \oplus \mathbf{F}_b \oplus \mathbf{F}_a - \mathbf{F}_b])\right),
\end{equation}
where $\oplus$ denotes concatenation, $\sigma$ is the sigmoid function, and the difference term captures relative quality variations.

\subsection{Learning Objectives}

We optimize TuningIQA using a multi-task learning framework with carefully designed loss functions.

\noindent\textbf{Confidence-Weighted Regression Loss.} To handle annotation uncertainty in fine-grained assessment, we design variance-based confidence weighting:
\begin{equation}
    \mathcal{L}_{reg} = \frac{1}{|\mathcal{D}|}\sum_{(I_i, s_i) \in \mathcal{D}} \exp(-v_i) \sum_{\text{attr}}|\hat{s}^{\text{attr}}_i - s^{\text{attr}}_i|,
\end{equation}
where $v_i$ denotes the annotation variance, adaptively reducing the impact of unreliable quality scores.

\noindent\textbf{Fine-grained Ranking Loss.} For rank learning, we employ the binary cross-entropy:
\begin{equation}
\mathcal{L}_{rank} = -\frac{1}{|\mathcal{P}|}\sum_{(I_p,I_q) \in \mathcal{P}} \sum_{\text{attr}} {BCE}(c^{\text{attr}}_{pq}, \hat{c}^{\text{attr}}_{pq}),
\end{equation}
where ${BCE}(c, \hat{c})$ denotes the binary cross-entropy function. The combined objective: $\mathcal{L} = \lambda_{reg}\mathcal{L}_{reg} + \lambda_{rank}\mathcal{L}_{rank}$ with empirically determined weights $\lambda_{reg}:\lambda_{rank} = 1:2$.

\begin{table}[]
\centering
\caption{Multi-attribute performance comparison.}
\label{tab:multi-att}
\small
\renewcommand{\arraystretch}{1.0}
\begin{tabular*}{\linewidth}{@{\extracolsep{\fill}}llcccc@{}}
\toprule
\textbf{Method} & \textbf{Metric} & \textbf{Sharp.} & \textbf{Noise} & \textbf{Exp.} & \textbf{Face} \\
\midrule
CLIP-IQA & SRCC & 0.532 & 0.472 & 0.199 & 0.510 \\
& PLCC & 0.528 & 0.434 & 0.208 & 0.512 \\
& FG-ACC & 0.649 & 0.622 & 0.571 & 0.518 \\
\midrule
MT-A & SRCC & 0.896 & 0.804 & 0.782 & 0.807 \\
& PLCC & 0.943 & 0.898 & 0.863 & 0.802 \\
& FG-ACC & 0.688 & 0.665 & 0.621 & 0.582 \\
\midrule
SARQUE & SRCC & 0.868 & 0.743 & 0.700 & 0.754 \\
& PLCC & 0.930 & 0.876 & 0.804 & 0.754 \\
& FG-ACC & 0.666 & 0.645 & 0.602 & 0.547 \\
\midrule
\rowcolor[HTML]{EFEFEF}
TuningIQA & SRCC & \textbf{0.935} & \textbf{0.805} & \textbf{0.897} & \textbf{0.885} \\
\rowcolor[HTML]{EFEFEF}
& PLCC & \textbf{0.963} & \textbf{0.905} & \textbf{0.949} & \textbf{0.880} \\
\rowcolor[HTML]{EFEFEF}
& FG-ACC & \textbf{0.773} & \textbf{0.746} & \textbf{0.776} & \textbf{0.725} \\
\bottomrule
\end{tabular*}
\end{table}

\section{Experiments}

\subsection{Performance Evaluation on FGLive-10K}

\textbf{Evaluation Protocol.} We compare the proposed TuningIQA model 
 against handcrafted and state-of-the-art learning-based BIQA models. All learning-based models are trained on FGLive-10K following strict scene-level division. We evaluate using SRCC, PLCC for score prediction, and fine-grained accuracy for pairwise ranking.

\noindent\textbf{Implementation Details.} We implement TuningIQA using PyTorch and train on NVIDIA RTX4090 GPUs. The model is optimized end-to-end using AdamW \cite{loshchilov2018decoupled} with cosine annealing \cite{loshchilov2017sgdr} at a maximum learning rate of $1 \times 10^{-5}$ and batch size of 64. Input images are resized to $256 \times 256$ and randomly cropped to $224 \times 224$ during training. Standard augmentations (horizontal/vertical flipping) are applied while avoiding quality-distorting operations. The model is trained for 5 epochs on all image pairs from the training set. We implement TuningIQA based on lightweight backbones to ensure application efficiency.

\textbf{Overall Performance.} Table \ref{tab:main} summarizes the performance comparison between TuningIQA and state-of-the-art BIQA methods on the FGLive-10K dataset. It is easily observed that while existing methods achieve reasonable score regression performance, they exhibit significantly poor fine-grained ranking capabilities, highlighting the inadequacy of score-based methods for subtle quality discrimination, which is essential in camera tuning. In contrast, TuningIQA achieves superior performance in both tasks, demonstrating the effectiveness of joint regression and ranking learning.

\textbf{Multi-attribute Evaluation.} To evaluate multi-attribute assessment capabilities, we further compare TuningIQA with three SOTA methods that support quality attribute prediction: CLIP-IQA, MT-A, and SARQUE. As shown in Table \ref{tab:multi-att}, face quality assessment proves particularly challenging for existing methods, where TuningIQA achieves superior performance through human-aware feature extraction across all attributes, demonstrating the effectiveness of our human-centric design.

\begin{figure}[t]
  \centering
  \includegraphics[width=0.847\linewidth]{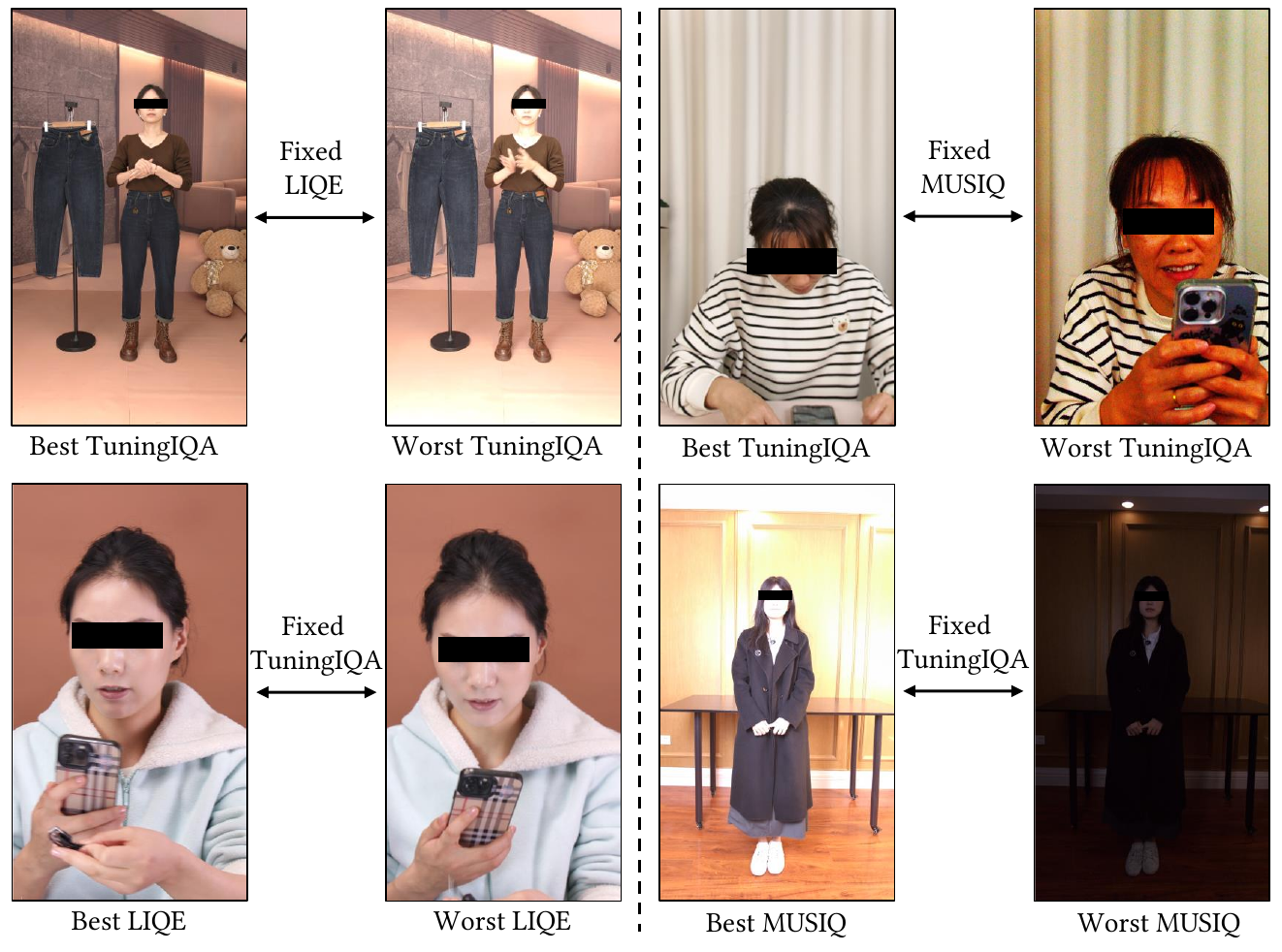}
  \caption{gMAD competition results on the FGLive-10K against LIQE and MUSIQ.}
  \label{fig:gmad}
\end{figure}

\textbf{Qualitative Analysis.} Figure \ref{fig:gmad} shows gMAD competition results \cite{gMAD}, exposing perceptual inconsistencies among models. LIQE and MUSIQ produce similar predictions for significantly different images or exaggerate minimal differences, while TuningIQA maintains correct discrimination, aligning better with human visual sensitivity.

\begin{table}[!t]
\centering
\caption{Ablation study. Results report average performance across overall score and attributes.}
\label{tab:ablation}
\small
\renewcommand{\arraystretch}{1.0}
\setlength{\tabcolsep}{5pt}
\begin{tabular*}{\linewidth}{@{\extracolsep{\fill}}lcccc@{}}
\toprule
\textbf{Dataset} & \textbf{Config.} & \textbf{SRCC↑} & \textbf{PLCC↑} & \textbf{FG-ACC↑} \\
\midrule
\multirow{2}{*}{FGLive-10K} & Baseline & 0.8757 & 0.9201 & 0.7258 \\
& w/ HFE & \textbf{0.8917} & \textbf{0.9267} & \textbf{0.7496} \\
\midrule
\multirow{3}{*}{FGLive-p} & Baseline & 0.8206 & 0.8768 & 0.7092 \\
& w/ GCPF & 0.8552 & 0.8963 & 0.7308 \\
& w/ GCPF+HFE & \textbf{0.8613} & \textbf{0.9009} & \textbf{0.7382} \\
\bottomrule
\end{tabular*}
\end{table}

\textbf{Ablation Study.} Table \ref{tab:ablation} presents ablation study results on two datasets: FGLive-10K and FGLive-p (subset with camera parameters). We systematically analyze the contribution of the Human-aware Feature Extraction (HFE) module and Graph-based Camera Parameter Fusion (GCPF) module. The results show that HFE consistently improves performance across all metrics on FGLive-10K, with notable gains in FG-ACC. On the FGLive-p subset, GCPF demonstrates substantial improvements, validating the importance of modeling parameter interdependencies. The combination of both modules achieves the best performance, confirming their complementary effectiveness.

\begin{figure}[]
  \centering
  \includegraphics[width=0.95\linewidth]{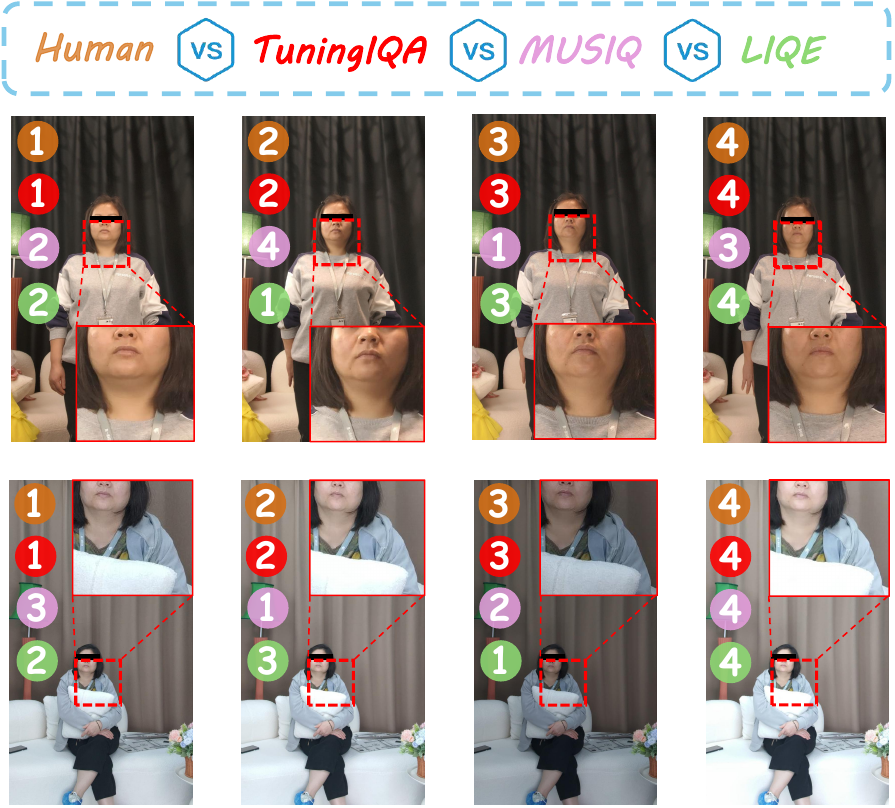}
  \caption{Qualitative comparison of IQA-guided livestreaming camera tuning results (best viewed zoomed in). }
  \label{fig:tuning_visual}
\end{figure}

\subsection{Application for Camera Tuning}

To validate the practical effectiveness of IQA metrics in guiding livestreaming camera parameter tuning, we simulate fine-grained parameter adjustment scenarios by incrementally adjusting camera parameters with minimal step sizes. We capture images after each adjustment and employ different metrics to rank the quality of these images, thereby identifying optimal parameters. TuningIQA performs ranking based on pairwise comparisons, while other methods rely on score-based ranking. As illustrated in Figure \ref{fig:tuning_visual}, we demonstrate the fine-grained quality differences resulting from small adjustments to Exposure Value and ISO, along with the ranking results from different methods. The results reveal the inaccuracy of score-based methods in fine-grained evaluation and their insufficient sensitivity to facial region quality. Additionally, we conduct a subjective experiment to compare the top-1 image quality between TuningIQA and LIQE/MUSIQ ranking results. Evaluation across 44 diverse scenarios with 12 volunteers shows that TuningIQA achieves \textbf{76\%} and \textbf{74\%} win rates against LIQE and MUSIQ respectively. These significant improvements validate that fine-grained quality assessment capabilities are essential for practical camera tuning applications, where subtle parameter adjustments can substantially impact user experience despite minimal changes in overall image appearance.

\section{Conclusion}

This work presents FGLive-10K dataset and TuningIQA, advancing fine-grained BIQA for livestreaming camera tuning. Our findings reveal that existing score-based BIQA methods, while achieving reasonable score regression, fundamentally struggle with fine-grained discrimination essential for camera optimization. Through human-aware feature extraction and graph-based parameter fusion, TuningIQA jointly model multi-attribute regression and fine-grained ranking. Extensive experiments demonstrate TuningIQA’s superior performance over existing methods and its practical effectiveness in providing actionable guidance for automated camera parameter optimization. 

\bibliography{aaai2026}

\end{document}